\documentstyle[epsfig,graphics]{mn}

\begin{document}

\newcommand{\Zsolar}{\mbox{\,$\rm Z_{\odot}$}}
\newcommand{\etal}{{et al.}\ }
\newcommand{\ang}{\mbox{$\rm \AA$}}
\newcommand{\xs}{$\chi^{2}$}

\title[The Sun, stellar-population models, and the age estimation of high-redshift 
galaxies.]{The Sun, stellar-population models, and the age estimation of
high-redshift galaxies}
\author[L.A. Nolan \etal.]
{L.A. Nolan, J.S. Dunlop $\&$  R. Jimenez.
\\
Institute for Astronomy, University of Edinburgh, Royal Observatory, Edinburgh, EH9~3HJ}

\date{Submitted for publication in MNRAS}

\maketitle
  
\begin{abstract}

Given sufficiently deep optical spectroscopy, the age estimation of high-redshift ($z > 1$) galaxies has been claimed to be a
relatively robust process ({\it e.g.} Dunlop et al.
1996) due to the fact that, for ages $< 5$Gyr, the
near-ultraviolet light of a stellar population is expected to be
dominated by `well-understood' main-sequence (MS) stars.
Recently, however, the reliability of this process has been called into 
question by Yi \etal (2000), who claim to have developed models in which 
the spectrum produced by the main sequence reddens much more rapidly than 
in the models of Jimenez \etal (2000a), leading to much younger age estimates for 
the reddest known high-redshift ellipticals. In support of their revised age 
estimates, Yi \etal cite the fact that their models can reproduce the 
spectrum of the Sun at an age of 5 Gyr, whereas the solar spectrum is not 
reproduced by the Jimenez \etal models until $\simeq 10$ Gyr. Here we confirm
this discrepancy, but point out 
that this is in fact a {\it strength} of the Jimenez \etal models and indicative 
of some flaw in the models of Yi \etal (which, in effect, imply 
that the Sun will turn into a red giant any minute now). We have also 
explored 
the models of Worthey (1994) (which are known to differ greatly from those 
of Jimenez \etal in the treatment of post-MS evolution) and find that the 
main-sequence component of Worthey's models also cannot reproduce the solar 
spectrum until an age of 9-10 Gyr. We conclude that either the models of 
Yi \etal are not as main-sequence dominated at 4-5 Gyr as claimed, or 
that the stellar evolutionary timescale in these models is in error by a 
factor possibly as high as two. Our current best estimate of the 
age of the oldest
galaxies at $z \simeq 1.5$ thus remains $3-4$ Gyr, as we confirm with
a new analysis of the existing data using the updated solar-metallicity models of both
Jimenez \etal and Worthey.
Finally, by fitting a mixed metallicity model to the Sun, we
demonstrate that, given rest-frame ultraviolet data of sufficient quality, 
it should be possible to break the age-metallicity degeneracy when
analyzing the spectra of high-redshift galaxies.

\end{abstract}


\section{Introduction}
For over a decade now, astronomers have attempted to estimate 
the ages of high-redshift galaxies using broad-band optical-infrared photometry
({\it e.g.} Lilly 1988; Dunlop et al. 1989, Chambers \& Charlot 1990).
Unfortunately, however, the derived ages have been rendered virtually
meaningless by disagreements between modellers over post main-sequence
evolution ({\it e.g.} Charlot, Worthey \& Bressan 1996), and by the
extreme susceptibility of such relatively crude broad-band 
data to dust reddening, emission-line contamination etc.

In contrast, it has long been anticipated that relatively robust age
constraints for high-redshift galaxies could be derived given rest-frame 
near-ultraviolet spectra of sufficient quality. This is because, for the 
potential ages of interest at $z > 1$ (i.e. ages $<$ 5 Gyr), the ultraviolet
light of a stellar population is expected to be dominated by stars close 
to the turn-off point of the `well-understood' main sequence (MS) ({\it
e.g.} Magris \& Bruzual 1993). 

With the advent of deep optical spectroscopy on 10-m class telescopes, it has 
now proved possible to put this technique into practice. In
particular, Dunlop et al. (1996) were able
to use a deep Keck spectrum of the $z = 1.5$ radio galaxy LBDS 53W091 to
first confirm that its near-ultraviolet spectrum was indeed dominated by
starlight, and then to extract an age constraint of $ > 3$Gyr based
primarily on comparison with a main-sequence only model of an evolving stellar
population. Spinrad et al. (1997) explored further the reliability of
this age estimate, and confirmed that the best agreement between 
ages derived using alternative evolutionary synthesis models was obtained
if fitting was confined to the detailed shape of the near-ultraviolet 
spectral energy distribution.

Not surprisingly, given its implications for cosmology (for $H_0 = 70
{\rm km s^{-1} Mpc^{-1}}$, the age of an Einstein-de Sitter universe at $z
= 1.5$ is only 2.3 Gyr), this result has been the subject of subsequent
close scrutiny, and claims that 53W091 is in fact less than 2 Gyr old
have been put forward by, for example,  Bruzual \& Magris (1997).
However, Dunlop (1999) has argued that such young ages are only deduced
using some models if the near-infrared photometry is also included in the fitting process,
once again placing undesirable emphasis on the reliability of the modelling of post
main-sequence evolution (a point previously also explored by Spinrad et
al. 1997). Moreover, Dunlop (1999) has shown that, certainly for the
slightly redder $z = 1.43$ galaxy 53W069, if fitting is
confined to the Keck spectroscopic data (Dey et al. 2000), the models of Bruzual \& Charlot
(1993), Worthey (1994), and Jimenez et al. (2000a) {\it all} lead to the
conclusion that its stellar population is $> 3$ Gyr old (assuming solar
metallicity).

Most recently, however, the reliability of even this near-ultraviolet
spectroscopic age-dating has been called into question by Yi et al. (2000).
Yi \etal (2000), claim to have derived a much younger age for 53W091, but
also claim that this age is not due to differences in post-MS evolution,
but rather to the fact that the spectrum produced by the main sequence 
in their models reddens much more rapidly than 
in the models of Jimenez \etal (2000a).
In support of their revised age 
estimates, Yi \etal cite the fact that their models can reproduce the 
spectrum of the Sun at an age of 5 Gyr, whereas the solar spectrum is not 
reproduced by the Jimenez \etal models until 8-10 Gyr.
It is unclear to us why a stellar population should be expected to mimic
the spectrum of the Sun at its current age ($\simeq 5$ Gyr); 
even if the light from the stellar population is dominated by stars 
near the main-sequence turnoff the Sun is not expected to leave the 
main sequence until an age of $\simeq 10$ Gyr (Jorgensen 1991).
Nevertheless this 
claim has motivated us to explicitly check the calibration of the
main-sequence evolution in alternative evolutionary
synthesis models of galaxy evolution. 

This is the main subject of the present paper. What we have done is to
take the 3 alternative and independent 
models of galaxy evolution developed by Yi et al.
(2000), Jimenez et al. (2000a) and Worthey (1994), and to check how
rapidly they evolve to mimic the solar spectrum with and (more importantly)
without inclusion of their post-MS components. 
The models are summarized in section 2, and the results of comparison
with the solar near-ultraviolet spectrum are presented in section 3. We
then proceed, in section 4, to use these models (again with and without post-MS
components) to check explicitly the extent to which the age estimates of
53W091 and 53W069 really are affected by different approaches to modelling
post-MS evolution. The main remaining uncertainty is the impact of having to
assume a value for the metallicity of the stellar population, and in
section 5 we explore whether, given near-ultraviolet data of sufficient quality, it may be
possible to break the well-known age-metallicity degeneracy. Finally,
our conclusions are summarized in section 6.

\section{The models}

\begin{figure*}
\centerline{\epsfig{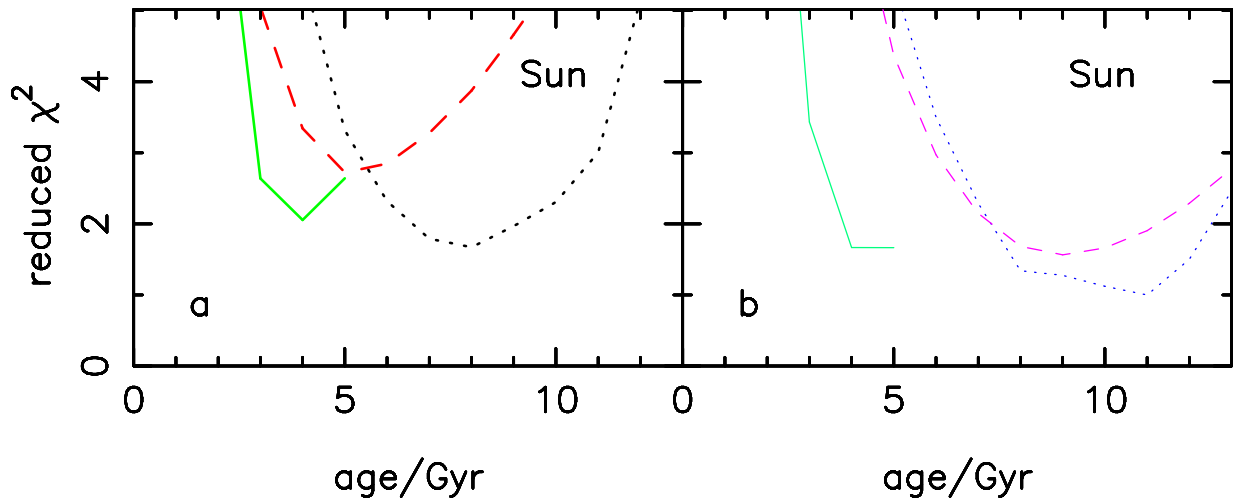}}
        
\caption{Reduced $\chi^2$ as a function of age for the six
alternative solar metallicity stellar population models fitted to the solar spectrum. 
On the left-hand panel are the results for the full stellar population models, 
and on the right-hand side are those for Worthey and Jimenez \etal main sequence (MS) 
only models and Yi \etal main sequence plus red giant branch (MSGB) models. 
Solid lines - Yi \etal (2000): dashed lines - Worthey (1994): dotted lines - 
Jimenez \etal (2000a). The MS / MSGB models result in a better fit than the full models. 
The best-fit age of Yi \etal's MSGB models differ from the best-fit ages of both Worthey's 
and Jimenez \etal's MS only models by a factor of order two, implying a
MS-turnoff age for the Sun of only 4-5 Gyr, compared with 9 or 11 Gyr
implied by the MS-only models of Worthey and Jimenez \etal respectively.}
\end{figure*}

\subsection{The models}

This study was motivated by the apparent disagreement reported by Yi et
al. (2000) between their own models and those of Jimenez et al. (2000a).
We are able to perform our own comparison of these models because Sukyoung
Yi has kindly supplied us with his model SEDs up to an age of 5 Gyr (Yi, private
communication). We also wanted to compare the predictions of a third
independent set of evolutionary synthesis models, and have chosen the
models of Worthey (1994) for this purpose. The reason for this choice was
that we already knew, from our previous modelling of 53W091 (Dunlop et al.
1996, Spinrad et al. 1997) that the models of Worthey (1994) also appear
to yield younger ages than those of Jimenez et al. (2000a), but for
reasons which we suspected were primarily due to a different treatment of
post-MS evolution (see Charlot et al. 1996).

Since the primary objective in this paper is to check the calibration of MS
evolution using the Sun, we have confined our attention only to models
which assume solar metallicity. However, we {\it are} interested in
removing any potential confusion introduced by different treatments of
post-MS evolution, because it is hard to be sure that the UV spectrum
of an instantaneous starburst really is completely dominated by MS stars
in all alternative models beyond an age of 2-3 Gyr (and beyond 5 Gyr it
is not expected to be). We have therefore
constructed a MS-only version of the models of Jimenez et al. (2000a), and
have also been supplied with a MS-only version of the models of Worthey
(1994) (Worthey, private communication). In both the Jimenez and the
Worthey models the isochrones have been cut off at the same point
(corresponding to point 55 in the Vandenberg grid in the case of
Worthey's models). A pure MS-only version of the Yi
et al. models was not available to us, but we do have access to a
main-sequence + giant branch (MSGB)
version, which is stripped of Horizontal Branch and Asymptotic Giant
Branch contributions, and should be an excellent approximation to an
MS-only model, at least in the near-ultraviolet for ages $< $ 5 Gyr, 
as Yi et al. themselves claim.

\subsection{$\chi^2$ minimization}

Since our main aim in this work was to calibrate the age-dating
of distant stellar populations based on the rest-frame near-ultraviolet
spectra, fitting was deliberately restricted to the spectral range $2000 -
4000$\AA. The best fit was determined by binning the data to the same
spectral resolution as the model in question, and then varying the age
and normalization as free parameters until $\chi^2$ was minimized.

For the high-redshift galaxies 53W091 and 53W069, the error on each
binned spectral datapoint was derived from propagation of the original
errors in the Keck optical spectra (see Dunlop et al. 1996; Dey et al. 2000). 
In the case of
the Sun, we are using the theoretical spectrum of Kurucz as the best
available representation of the true solar SED, and 
have simply assumed a constant flux-density error (i.e.
independent of wavelength), adjusted in size until 
reduced chi-squared ($\chi^2_{\nu}$) equalled unity for the very best
fitting model. It is thus only possible to compare the relative (rather
than absolute) ability of the different models to reproduce the solar
ultraviolet spectrum as the age of each model is varied.

\section{Comparison with the solar spectrum}

\subsection{Full models}

In Figure 1a we show reduced \xs as a function of age for each of the 
full stellar population models when age is varied in an attempt to best
reproduce the solar spectrum. The models of Yi \etal predict the youngest age, 
indicating that the near-ultraviolet spectrum produced by these models
best mimics that of the sun after an age of only 4 Gyr. 
The models of Worthey yield a best-fit age of 5 Gyr, while those of 
Jimenez \etal predict an age of 8 Gyr. From this plot it might appear that 
it is the models of Jimenez \etal that are most unusual, but it is important to
note that i)  it is the models of Jimenez \etal which yield the best
quality of fit to the solar spectrum, and ii) it is to be expected 
that the full stellar population models will {\it underestimate} the main-sequence 
turn-off (MSTO) age of the sun, because of the inclusion of post-MS
stars. Moreover, from this plot it is completely unclear how much of the
(substantial) difference between the derived best-fit ages can be
attributed to differing contributions of post-MS
stars to the galaxy ultraviolet SEDs.

\subsection{Main-sequence only models}

\begin{figure*}
\centerline{\epsfig{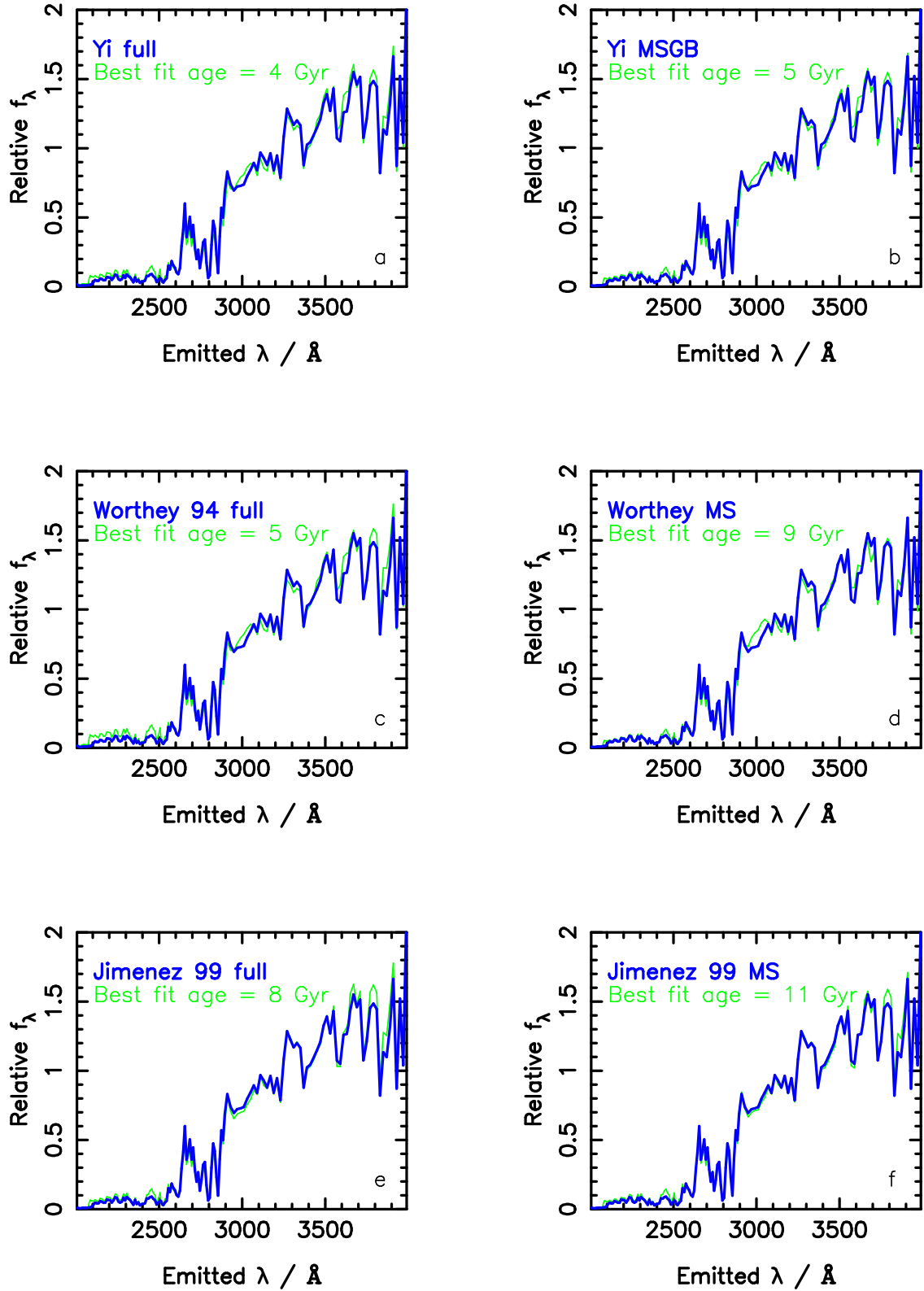}}

\caption{Best fits to the solar spectrum (black lines) for the various solar metallicity models (grey lines).  }
\end{figure*}

For a meaningful comparison between galaxy synthesis models and the
expected MS turn-off age of the sun, we really require to ensure that the
ultraviolet spectra produced by the models are completely MS dominated.
Therefore, in  Figure 1b we show reduced $\chi^2$ as a function 
of age in an analogous way to Figure 1a, but this time using models,
stripped as far as possible of post-MS contributions. This allows not
only a sensible comparison with the Sun, but also makes it possible to
assess the relative impact of post-MS contributions to the full model fits
shown in Figure 1a.

\begin{figure*}
\centerline{\epsfig{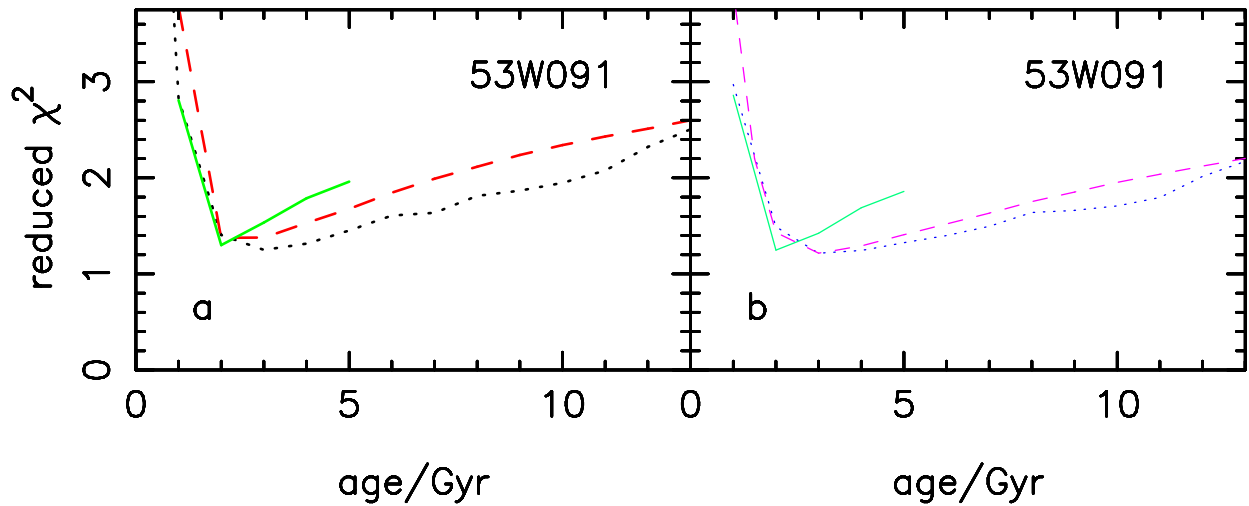}}

\caption{Reduced $\chi^2$ as a function of age for the six
alternative solar metallicity stellar population models fitted to the
near-ultraviolet spectrum of the $z = 1.55$ radio galaxy 53W091 (Dunlop
et al. 1996; Spinrad et al. 1997).
On the left-hand panel are the results for the full stellar population models, 
and on the right-hand side are those for Worthey and Jimenez \etal main sequence (MS) 
only models and Yi \etal main sequence plus red giant branch (MSGB) models. 
Solid lines - Yi \etal (2000): dashed lines - Worthey (1994): dotted lines - 
Jimenez \etal (2000a).}
\end{figure*}

This plot makes it clear that it is the models of Yi et al. that are
unusual, and apparently in error by a factor of two 
in terms of rate of MS evolution. Stripped of HB and AGB
contributions the predictions of the Yi et al. models are little changed,
indicating a best-fit age of 4-5 Gyr. In contrast, stripped of post-MS
evolution the models of Worthey and Jimenez \etal appear to be in good
agreement not only with each other (9 Gyr and 11 Gyr respectively) but 
also with the generally accepted MS turnoff age of the Sun (10.5 Gyr -
Jorgensen 1991).

It is hard to assess the likely impact on the predictions of the Yi et
al. models if the GB was also removed, but the lack of any dramatic change upon
removal of HB and AGB contributions does tend to support their own claim
that their models are already highly MS dominated at ages $<$ 5 Gyr. In
contrast the contribution of post-MS stars in the models of Worthey must
be relatively strong, even at 5 Gyr, because removal of HB + AGB moves
the derived age from 5 to 7 Gyr, and subsequent removal of remaining
post-MS stars completes the shift to 9 Gyr as shown in Figure 1b. This
therefore backs up the suggestion made in Jimenez et al. (2000a) that the
main difference between the models of Jimenez et al. and Worthey lies in
the strength of the AGB and RGB, but that the MS evolution in
both models is very similar, and yields sensible values for the turn-off
age of the Sun.

These results are tabulated in Table 1, and in Figure 2 we show
the best-fit model spectra superimposed on the solar
spectrum. It is worth noting (from both Figure 2 and Figure 1)
that, as one would hope, the MS-only models
in every case yield a better {\it quality} of fit to the solar spectrum
than do the full models which include some contribution from more evolved
stars. Interestingly, the MS-only models of Jimenez \etal yield 
both the best overall fit in terms of reduced $\chi^2$, and a MS-turnoff
age which most closely matches the accepted value of 10.5 Gyr (Jorgensen
1991).

In summary, either the timescale of MS evolution in the models of Yi et
al. is too short by a factor $\simeq 2$, or the GB contribution to the UV
spectrum at an age of 4 Gyr is unexpectedly large, in which case these
models would again be extremely unusual and the 
comparison with the age of the Sun as suggested by Yi et al. must
inevitably be meaningless.

\section{Comparison with the spectra of red galaxies at $z \simeq 1.5$} 

\subsection{LBDS 53W091}

The age implications of the deep Keck optical spectrum of the red mJy radio galaxy 53W091
have been previously discussed in some detail by Dunlop et al. (1996) and
Spinrad et al. (1997). However it is interesting and important to revisit
the age-determination of this object for a number of reasons.
First, the models of Jimenez \etal have been updated in the
intervening years. Second, the Yi et al. models did not exist in
1996/1997. Third, we have only recently obtained the MS-only versions of the
Worthey models. We also felt it was important to re-analyze this spectrum
given the claims made by Bruzual \& Margris (1997) and Yi et al. (2000)
that the most recent models yield best-fit ages for the stellar
population in 53W091 of less than 2 Gyr.

\begin{figure*}
\centerline{\epsfig{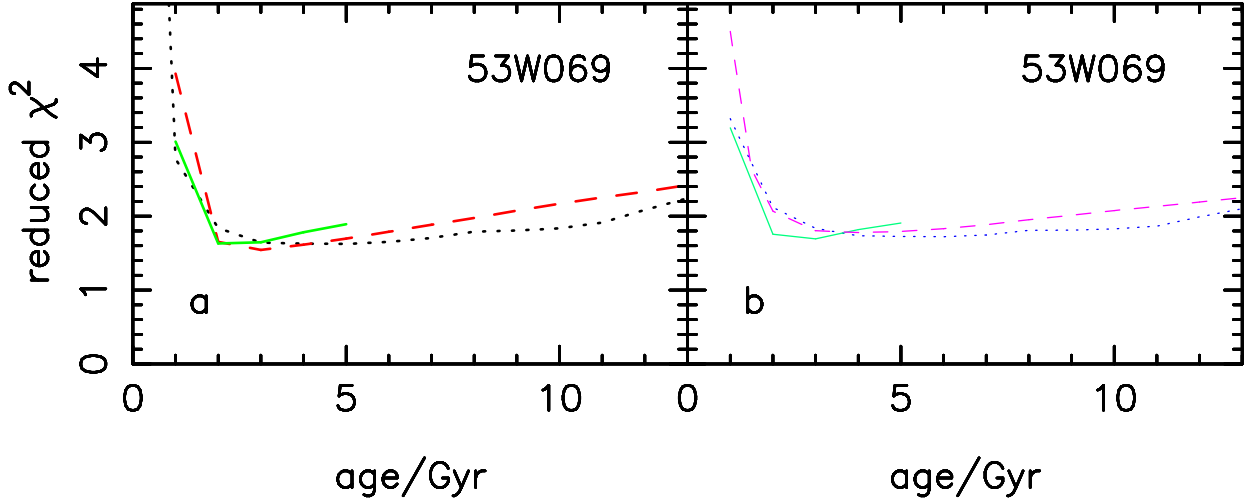}}

\caption{Reduced $\chi^2$ as a function of age for the six
alternative solar metallicity stellar population models fitted to the
near-ultraviolet spectrum of the $z = 1.43$ radio galaxy 53W069 (Dunlop
1999; Dey et al. 2000).
On the left-hand panel are the results for the full stellar population models, 
and on the right-hand side are those for Worthey and Jimenez \etal main sequence (MS) 
only models and Yi \etal main sequence plus red giant branch (MSGB) models. 
Solid lines - Yi \etal (2000): dashed lines - Worthey (1994): dotted lines - 
Jimenez \etal (2000a).}
\end{figure*}

Therefore in Figure 3 we show the results of fitting the same six models
(as fitted to the Sun above), to the rest-frame near-ultraviolet
SED of 53W091. The results, summarized in Table 1, are that both the
Jimenez et al. (2000a) models (full and MS only) yield a best-fit age of 3
Gyr (with ages as young as 2 Gyr formally excluded), both the Yi et al.
models (full and MSGB) yield a best fit age of 2 Gyr, while the best-fit
age yielded by the Worthey models changes from 2 Gyr to 3 Gyr if post-MS
contributions are excluded. Given the fact that we have already shown
that the MS clock in the Yi et al. models appears to be running up to a factor of 2 too
fast, this means that the argument over whether the age of 53W091 can or cannot be
younger than 3 Gyr comes down to a debate over the validity of the stronger AGB
and GB contributions in the Worthey models compared with the more
MS-dominated Jimenez \etal models. Jimenez et al. (2000a) argue that their
models include a better treatment of mass-loss in evolved stars which
results in less luminous RGB and AGB contributions.
In summary, the available evidence still appears to favour a minimum age
of 3 Gyr for this galaxy, subject to remaining uncertaintities over the
impact of possible non-solar metallicity (see section 5).

\subsection{LBDS 53W069}

As discussed by Dunlop (1999), the Keck spectrum of the even redder mJy radio galaxy 
53W069 ($z = 1.43$; Dey et al. 2000) appears to
offer the best example discovered to date of a highly-evolved coeval
stellar population at a redshift as high as $z \simeq 1.5$. Dunlop (1999)
found, from a comparison of their near-ultraviolet SEDs, that 53W069 is
significantly redder than 53W091, and that the SED of the latter galaxy 
can be decomposed into that of 53W069 plus a low-level blue
component which is approximately flat in $f_{\lambda}$.
It is therefore to be expected that model-fitting to the near-ultraviolet
SED of 53W069 might yield even older age limits than those derived above
for 53W091. 

In Figure 4 we show the results of fitting the same six
models as before to the near-ultraviolet SED of 53W069. The results can
again be found in tabulated form in Table 1. In summary, for this object
5 out of the 6 models yield a minimum age $ >  3$ Gyr, with only the Yi
et al. full models allowing an age as young as 2 Gyr. Since these models
appear flawed based on the solar comparison discussed in section 3, 
we conclude that it is extremely hard to escape the conclusion
that 53W069 is at least $3$ Gyr old. For this object the 
debate over whether the post-MS treatment of Jimenez et al. (2000a) is
to be preferred over that of Worthey (1994) translates into
an uncertainty over whether the best-fit age is in fact $> 3$ or $> 4$ Gyr
(see Table 1). 

For $H_0 = 70 {\rm km s^{-1} Mpc^{-1}}$, the age
of an Einstein-de Sitter Universe at $z = 1.43$ is only 2.5 Gyr.
Therefore, given the apparently flawed nature of the Yi et al. models, we
conclude that the only way that the age of 53W069 at $z = 1.43$ can be
contained within an Einstein-de Sitter Universe is if varying the assumed
metallicity from solar can in fact reduce the best-fit age to $< 2.5$ Gyr.

\begin{table*}

\caption{A summary of the best fit ages produced by fitting the 6
alternative models discussed in the text to the near-ultraviolet spectral
energy distribution of i) the sun (see Figures 1 and 2), ii) the $z = 1.55$ galaxy 
53W091 (see Figure 3), and iii) the $z = 1.43$ galaxy 53W069 (see Figure
4). In the case of the sun, the result of fitting the mixed metallicity
model discussed in section 5 is also given (see Figure 5). The value of
reduced $\chi^2$ is also given in each case, although in the case of the
fits to the Sun, the values of reduced $\chi^2$ can only be used to judge
the relative quality of the alternative model fits.}

\begin{tabular}{rlcc}

\\

\hline

 \\

	\large{object} & \large{model} & \large{best fit age / Gyr} & \large{ reduced \xs}  	\\

 \\

\hline  

\\

SUN & J Full & 8 & 1.67 \\
    & J MS   &11 & 1.00	\\
    & W Full & 5 & 2.72 \\
    & W MS   & 9 & 1.57	\\
    & Y Full & 4 & 2.06 \\
    & Y MSGB & 4 & 1.67	\\
    & J 3Z   & 8 & 1.37 \\

\\
     
\hline  

\\

LBDS 53W091 & J Full & 3 & 1.25	\\
            & J MS   & 3 & 1.22	\\
            & W Full & 2 & 1.38	\\
            & W MS   & 3 & 1.22	\\
            & Y Full & 2 & 1.30	\\
            & Y MSGB & 2 & 1.25	\\

\\

\hline  

\\

LBDS 53W069 & J Full & 5 & 1.63	\\
            & J MS   & 6 & 1.72	\\
            & W Full & 3 & 1.54	\\
            & W MS   & 4 & 1.78	\\
            & Y Full & 2 & 1.63	\\
            & Y MSGB & 3 & 1.69	\\
\\

\hline

\end{tabular}

\end{table*}

\section{Breaking the age-metallicity degeneracy}

The potential severity of the age-metallicity degeneracy was highlighted 
in the work of Worthey (1994), and acknowledged by Dunlop et al. (1996)
in their original attempts to determine the age of 53W091. It is clear
that assuming a metallicity of twice solar for the entire stellar
population results in the lower limit to the derived age of both 53W091 and 53W069 
falling below 2.5 Gyr.

Spinrad et al. (1997) presented arguments that, even assuming these
galaxies possess a high-metallicity core, adoption of a universally high
metallicity is inappropriate when analyzing the integrated light 
from the central $\simeq 10$ kpc of an elliptical galaxy (as sampled at
$z > 1$ by the Keck
spectroscopic slit). Moreover, both Jimenez et al. (2000b) and Yi et al.
(2000) have now attempted to produce more realistic mixed metallicity models 
and have independently found that such mixed-metallicity models in fact
yield very similar ages to simple {\it solar} metallicity models (despite
the basic disagreement between model timescales detailed above). This
appears to be true, even when the average metallicity is twice solar,
simply because it is the low metallicity component which dominates the
light shortward of 3000\AA\ (Jimenez et al. 2000b).

Ideally, however, we would like to be able to fit both 
metallicity mix and age
to the data. The extent to which this is possible obviously depends both
on the quality of the spectroscopic data available, and on the presence
of (primarily) metallicity-dependent and age-dependent features within
the available spectral range.

This approach will be explored further in a separate paper, but here we
present evidence that the age metallicity degeneracy can, at least 
in principle, be
broken with data in the spectral range 2000\AA -- 4000\AA, using the Sun
as a test case.

A mixed-metallicity model was constructed from the 0.2\Zsolar, \Zsolar\ and 
2.5\Zsolar\ full models of Jimenez \etal, with the relative contributions of the 
different metallicity models allowed to vary as free parameters. 
The ability of this mixed-metallicity model to reproduce the metallicity and MSTO age of 
the sun is a test of the stellar population synthesis models' ability to break the 
age-metallicity degeneracy given data in this near-ultraviolet spectral
range.

The solar spectrum was rebinned in the same way as for the 
solar-metallicity only fitting process. The mixed-metallicity model flux was built from 
normalised SED's, so that

\begin{eqnarray}
	F_{3Z,\lambda,age} & = & X_{0.2{\tiny{\Zsolar}}} f_{0.2{\tiny{\Zsolar}},\lambda,age} + X_{{\tiny{\Zsolar}}} f_{{\tiny{\Zsolar}},\lambda,age} + \nonumber \\
                           &   & X_{2.5{\tiny{\Zsolar}}} f_{2.5{\tiny{\Zsolar}},\lambda,age} \nonumber
\end{eqnarray}

\noindent
where $F_{3Z,\lambda,age}$ is the 
mixed metallicity flux per unit wavelength in the bin centred on 
wavelength $\lambda$ at $age$ Gyr, $f_{{\tiny{Z}},\lambda,age}$ is the flux per 
unit wavelength in the bin centred on wavelength $\lambda$ of the model at  $age$ Gyr, 
and metallicity Z, and $X_{0.2{\tiny{Z}}}$ is the fractional contribution to $F_{3Z,age}$ 
by $f_{{\tiny{Z}},age}$.

A \xs fit was used to determine the best-fit age, total normalization, 
and values of $X_{0.2{\tiny{\Zsolar}}}$, $X_{{\tiny{\Zsolar}}}$ 
and $X_{2.5{\tiny{\Zsolar}}}$.

In Figure 5 we show the results of this mixed-metallicity fitting
procedure. Comparison of the left-hand plot with the dotted line in
Figure 1a demonstrates the extent to which allowing metallicity to vary
has weakened the constraint on age. However, the 
best-fit age is still 8 Gyr, as for the full solar-metallicity only models. 
Figure 5b shows the evolution of the fractional contributions to the flux of the 
different metallicity SED's. At the best-fit age of 8 Gyr, the flux is clearly 
completely dominated by the solar metallicity SED, as it should be. 
$X_{0.2{\tiny{\Zsolar}}}$ is 0.01, $X_{{\tiny{\Zsolar}}}$ is 0.99 and there is 
no contribution from the 2.5\Zsolar\ model. 
This gives a mean metallicity of 0.99\Zsolar. 
In summary, given data of this (obviously excellent) quality, the
mixed-metallicity model can return both the correct metallicity and the
correct age to high accuracy (where here the correct age does not mean
the turnoff age of the Sun, but rather the age of 8 Gyr which was
returned using the full model with the correct metallicity).

\begin{figure*}
\centerline{\epsfig{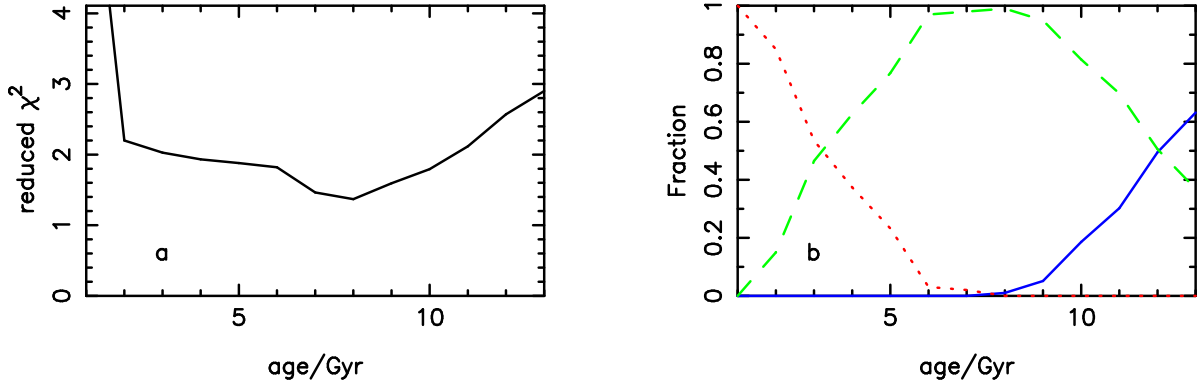}}
        
\caption{Left-hand side: reduced $\chi^2$ as a function of age 
for the mixed metallicity model (see \S 5 for details). The best-fit age
is 8 Gyr, just as for the full solar-metallicity models of Jimenez et al.
(2000). Right-hand side: fractional contributions to the mixed metallicity model 
of the different metallicity components as a function of age, 
{\em i.e.} 0.2\Zsolar\ (solid line), \Zsolar\ (dashed line) and 2.5\Zsolar\ (dotted line). 
The large contribution of the super-solar component at young ages, and
the large contribution of the sub-solar component at large ages reflects
the well-known age-metallicity degeneracy. However, at the best fit age of 8 Gyr, 
the mean metallicity is 0.99\Zsolar, with the solar metallicity
contribution completely dominating the model.}

\end{figure*}

\section{Conclusion}

In this paper we have not attempted a model-maker's comparison of
different evolutionary synthesis models, deliberately not entering into
the debate over which model uses the most trustworthy components, such
as isochrones, stellar atmospheres
etc. These issues will be addressed in a future paper (Jimenez et al., 2000c). Instead, we have addressed the simple issue of whether the
near-ultraviolet spectral energy distribution of the Sun is reproduced by
the main-sequence only components of three independent models at an age
commensurate with current estimates of the MS-turnoff age of the Sun.
This test was suggested by Yi et al. (2000), but in fact leads us to the
conclusion that the evolutionary timescale of the main sequence in the Yi
et al. models is anomalous when compared with the either the models of
Worthey (1994), Jimenez et al. (2000a) or the current best estimate of
the MS turn-off age of the Sun. 

We have also re-addressed the issue of the
extent to which varying contributions from post-MS phases of stellar
evolution can affect the age estimation of galaxies at $z \simeq 1.5$
from rest-frame ultraviolet spectra. We find that (assuming solar
metallicity) the minimum age of
the dominant stellar population in the the $z = 1.55$ radio galaxy 53W091 
is 2 Gyr if one invokes the relatively strong AGB/RGB components included
in the models of Worthey (1994), and 3 Gyr is one adopts the weaker
post-MS contributions included in the models of Jimenez et al. (2000a).
In the case of the even more passive $z = 1.43$ radio galaxy 53W069,
these numbers become 3 Gyr and 4 Gyr respectively. 

Finally, as stated above, these figures are derived on the assumption of
approximately solar metallicity. Yi et al. (2000) and Jimenez et
al. (2000b) have in fact shown that the age metallicity degeneracy in a
realistic mixed-metallicity population is probably not nearly as severe
as previously feared. Nevertheless, ideally it is clearly desireable to
determine both age and metallicity directly from the observations.
Using the spectrum of the Sun, we have shown that, given data of
sufficient quality, it should in principle be possible to break the
well-known age-metallicity degeneracy using data covering only the
near-ultraviolet spectral range which is accessible when observing
galaxies at $z > 1$ with optical spectrographs. The possibility of
breaking the age metallicity degeneracy will be explored further in a
subsequent paper.

\vspace*{1cm}

\noindent
{\bf ACKNOWLEDGEMENTS}\\
We would especially like to thank Sukyoung Yi for making his models
available to us in exchange for the Jimenez models, and also 
Guy Worthey for the willingness and
speed with which he supplied us with specific versions of his model
predictions, customised to meet our specific requirements. Louisa Nolan
acknowledges the support of a PPARC studentship, and Raul Jimenez
acknowledges the support of a PPARC Advanced Fellowship.

\vspace*{1cm}

\noindent
{\bf REFERENCES}\\
Bruzual, G., Charlot, S., 1993, ApJ, 405, 538\\
Bruzual, G.A., Magris, G.C., 1997, astro-ph/9707154\\
Chambers, K.C., Charlot, S., 1990, ApJ, 348, L1\\
Charlot, S., Worthey, G., Bressan, A., 1996, ApJ, 457, 626\\
Dey, A., et al. 2000, ApJ, in preparation\\
Dunlop, J.S., 1999, In: {\em `The Most Distant Radio Galaxies'}, KNAW
Colloguium Amsterdam, p.71, eds. Rottgering, H.J.A., Best, P., Lehnert, M.D., Kluwer\\
Dunlop, J.S., Guiderdoni, B., Rocca-Volmerange, B., Peacock, J.A.,
Longair, M.S., 1989, MNRAS, 240, 257\\
Dunlop, J., Peacock, J., Spinrad, H., Dey, A., Jimenez, R., Stern, D., Windhorst, R., 1996, Nature, 381, 581\\
Jimenez R., Dunlop J.S., Peacock J.A., Padoan P., MacDonald J.,
J$\o$rgensen U.G., 2000a, MNRAS, in press\\
Jimenez, R., Padoan, P., Juvela, M., Bowen, D.V., Dunlop, J.S.,
Matteucci, F., 2000b, ApJ, in press\\
Jimenez R. et al., 2000c, in preparation\\
Lilly, S.J., 1988, ApJ, 333, 161\\
Jorgensen, U.G., 1991, A\&A, 246, 118\\
Magris, G.C., Bruzual, G.B., 1993, ApJ, 417, 102\\
Spinrad, H., Dey, A., Stern, D., Dunlop, J.S., Peacock, J.,
Jimenez, R.,  Windhorst, R., 1997, ApJ, 484, 581\\
Worthey G., 1994, ApJS, 95, 107\\
Yi S., Brown T.M., Heap S., Hubeny I., Landsman W., Lanz T., Sweigart A.,
2000, ApJ, in press (astro-ph/9911067).\\

\end{document}